\documentclass[aps,prx,twocolumn,floatfix,longbibliography,superscriptaddress]{revtex4-2}

\usepackage{amsmath}
\usepackage{amssymb}
\usepackage{times}
\usepackage{braket}
\usepackage[pdftex]{graphicx}
\usepackage{color}
\usepackage{blindtext}
\usepackage{nicefrac}
\usepackage[colorlinks=true, citecolor=blue, urlcolor=blue, linkcolor=blue]{hyperref}

\usepackage{todonotes}

\renewcommand{\vec}[1]{\boldsymbol{#1}}

\begin{document}

\title{Exact theory of chirality-dependent p-wave magnetism and Edelstein effect in spin spirals}

\author{B{\"o}rge G{\"o}bel }
\email[Correspondence email address: ]{boerge.goebel@physik.uni-halle.de}
\affiliation{Institute of Physics and Halle-Berlin-Regensburg Cluster of Excellence CCE, Martin Luther University Halle-Wittenberg, 06120 Halle (Saale), Germany}

\author{Tom G. Saunderson}
\affiliation{Institute of Physics and Halle-Berlin-Regensburg Cluster of Excellence CCE, Martin Luther University Halle-Wittenberg, 06120 Halle (Saale), Germany}

\author{Freia Opfermann}
\affiliation{Institute of Physics and Halle-Berlin-Regensburg Cluster of Excellence CCE, Martin Luther University Halle-Wittenberg, 06120 Halle (Saale), Germany}

\author{Lennart Schimpf}
\affiliation{Institute of Physics and Halle-Berlin-Regensburg Cluster of Excellence CCE, Martin Luther University Halle-Wittenberg, 06120 Halle (Saale), Germany}

\author{Ersoy \c{S}a\c{s}{\i}o\u{g}lu}
\affiliation{Institute of Physics and Halle-Berlin-Regensburg Cluster of Excellence CCE, Martin Luther University Halle-Wittenberg, 06120 Halle (Saale), Germany}

\author{Samir Lounis}
\affiliation{Institute of Physics and Halle-Berlin-Regensburg Cluster of Excellence CCE, Martin Luther University Halle-Wittenberg, 06120 Halle (Saale), Germany}

\date{\today}

\begin{abstract}
Spin-momentum locking is widely regarded as a hallmark of relativistic spin-orbit coupling. Here we demonstrate analytically that it can instead emerge solely from magnetic chirality. Solving the minimal tight-binding model of electrons coupled to a spin spiral using a generalized Bloch theorem, we show that the spiral generates chirality-dependent p-wave magnetism characterized by the antisymmetric spin texture $\vec{s}(\vec{k})=-\vec{s}(-\vec{k})$. The exact solution further yields closed-form expressions for the electrical conductivity and the spin Edelstein susceptibility, revealing a microscopic mechanism by which magnetic chirality alone can generate spin polarization without spin-orbit coupling, with direct implications also for chirality-induced spin selectivity. In the strong exchange-coupling regime, the spin-dependent physics of the spin spiral becomes directly analogous to the orbital-dependent physics of electrons propagating through a non-magnetic helix. Our work establishes a minimal exactly solvable model of chirality-induced spin-momentum locking and spin-charge conversion beyond the conventional spin-orbit coupled paradigm.\end{abstract}

\maketitle


Momentum-dependent spin polarization has emerged as a central concept in condensed matter physics, underpinning a wide range of spintronic phenomena. Traditionally, such spin textures originate from relativistic spin-orbit coupling, as realized in topological materials~\cite{hasan2010colloquium} or Rashba systems~\cite{rashba1960properties,bychkov1984oscillatory}, where they cause a current-induced spin polarization labeled Edelstein effect~\cite{edelstein1990spin}. 
More recently, antiferromagnetic crystals have emerged as an alternative platform for generating momentum-dependent spin polarization. In particular, altermagnets~\cite{vsmejkal2022emerging,smejkal2022beyond,amin2024nanoscale,fedchenko2024observation} exhibit spin splitting without net magnetization, while the recently introduced concept of antialtermagnetism~\cite{hellenes2023p} extends this framework to odd-parity (p-wave) spin textures in reciprocal space that also give rise to an Edelstein effect~\cite{chakraborty2025highly}. These developments raise a natural question: Can p-wave magnetism arise in the complete absence of spin-orbit coupling, driven solely by magnetic texture?

A particularly appealing candidate is provided by spin spirals~\cite{yoshimori1959new}, the simplest realization of chiral noncollinear magnetism~\cite{gobel2021beyond}. When an electron propagates through a spin spiral, its spin continuously aligns with the non-collinear magnetic moments and consequently undergoes a coherent precession. This precession can be described by an effective magnetic field acting on the electron spin. Since the sense of the precession reverses upon changing the propagation direction, the effective field is odd under momentum inversion, giving rise to an antisymmetric spin texture as a function of the crystal momentum,
$\vec{s}(\vec{k})=-\vec{s}(-\vec{k})$,
corresponding to p-wave magnetism. Moreover, the sign of the effective field is determined by the chirality of the spin spiral, establishing a direct correspondence between magnetic handedness in real space and spin polarization in momentum space. This chirality-dependent spin splitting has recently been observed experimentally~\cite{song2025electrical,yamada2025metallic}, providing compelling evidence that noncollinear magnetic textures can generate momentum-space spin polarization without requiring intrinsic relativistic interactions. 
While effective models have recently been developed to describe generic p-wave magnets and their transport properties~\cite{brekke2024minimal}, a microscopic analytical theory demonstrating how p-wave magnetism emerges from a chiral magnetic texture is still lacking.

Here, we derive the microscopic origin of chirality-induced p-wave magnetism by solving exactly the minimal tight-binding model of electrons coupled to a spin spiral.
Employing a generalized Bloch theorem, we obtain closed-form expressions for the band structure and spin texture, revealing the emergence of chirality-dependent p-wave magnetism. Building on this exact solution, we derive analytical expressions for the electrical conductivity and the chirality-dependent spin Edelstein susceptibility, demonstrating that chirality-induced spin polarization arises entirely from the magnetic texture, without any spin-orbit coupling. We further identify the distinct transport regimes from weak to strong exchange coupling and uncover an intimate analogy to the orbital physics of electrons propagating through non-magnetic chiral structures~\cite{gobel2025chirality,gobel2025chirality2}. Our results establish a minimal exactly solvable model of chirality-induced p-wave magnetism and provide a transparent microscopic foundation for non-relativistic spin-momentum locking and spin-charge conversion generated solely by magnetic chirality.

\begin{figure*}[t!]
    \centering
    \includegraphics[width=\textwidth]{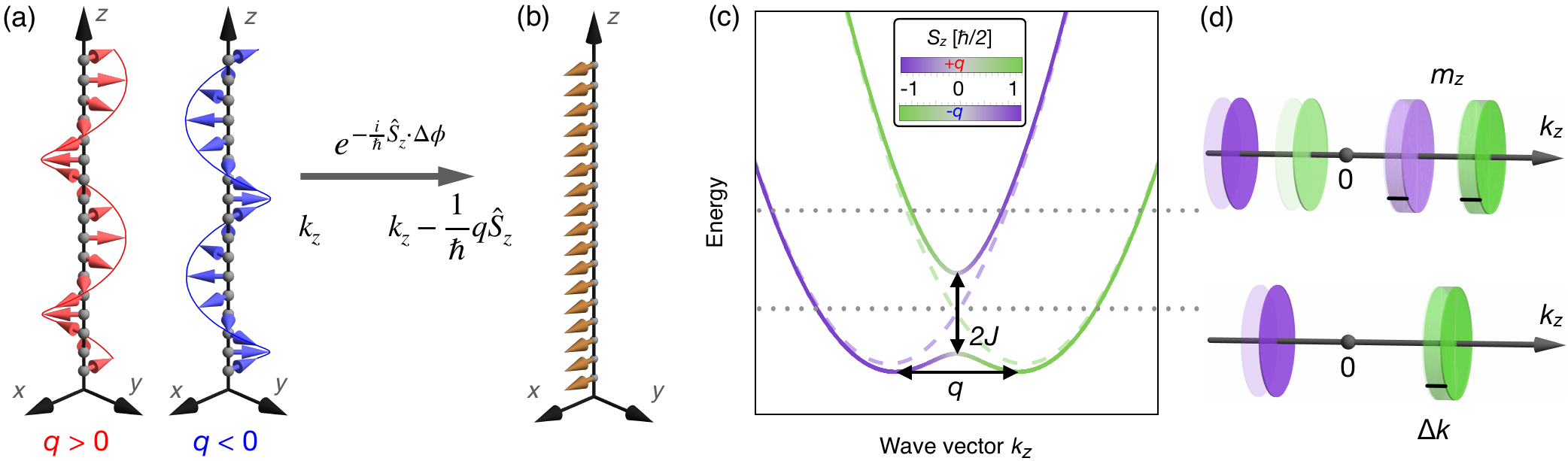}
    \caption{Schematic representation of emerging chirality-dependent p-wave magnetism and Edelstein effect based on generalized Bloch theorem. (a) Spin spirals with positive and negative chirality. (b) After transforming into the rotating spin frame, the spatially varying exchange field becomes a uniform ferromagnetic background, while the electronic states acquire a spin- and chirality-dependent momentum shift. (c) Dashed lines: Schematic band structure in the weak-coupling limit, illustrating the rigid momentum shifts of the spin-up and spin-down bands. Solid lines: hybridized bands in the presence of a finite exchange coupling $J$. The color scale indicates the expectation value of $s_z$. (d) Schematic illustration of the spin Edelstein effect for positive chirality ($q>0$). An applied electric field shifts the electronic distribution by $\Delta k$, generating a nonequilibrium spin magnetic moment $m_z$ proportional to the shaded regions. The upper and lower panels correspond to Fermi levels outside and inside the exchange gap indicated by dotted lines in (c), respectively. Inside the gap, only a single band contributes, resulting in a strongly enhanced Edelstein response, whereas outside the gap the contributions from the two bands partially cancel.}
    \label{fig:geometry}
\end{figure*}

\paragraph*{Generalized Bloch theorem.}
Throughout this paper, we consider a spin spiral with a wave vector pointing along $z$ and wave number $|q|=2\pi/\lambda$. The sign of $q$ controls the spiral's chirality; clockwise versus counter-clockwise [see Fig.~\ref{fig:geometry}(a)]. We consider a one-dimensional periodic description; the spiral has a period of $\lambda=Na$ and is formed by $N$ magnetic moments $\{\vec {m}_n\}_{n=1,\dots,N}$ (here unit vectors) per unit cell that are oriented perpendicularly to the propagation direction. $a$ is the lattice constant. 

Without the spin spiral, the wave function spinor $\psi^{\uparrow\downarrow}_k(z)$ is the product of a lattice periodic function, $u^{\uparrow\downarrow}_k(z+a)=u^{\uparrow\downarrow}_k(z)$, and a phase factor,
according to Bloch's theorem. Therefore, the wave function at two neighboring sites differs by a phase factor $e^{ika}$.
Once the spin spiral is considered, the spin at $z+a$ is rotated by $\phi=qa$ in the $xy$ plane with respect to the spin at $z$. The wave function does not fulfill the Bloch theorem as defined above anymore, because the spin-dependent potential, accounting for the spiral, is not periodic in $a$ anymore but in $\lambda$. The translational symmetry can nevertheless be restored by transforming to the rotating spin frame of reference. The corresponding unitary operator is a rotation matrix. For two neighboring moments it is $\hat{R}_z=e^{-i\phi\hat{s}_z/\hbar}$ with $\hat{s}_z=\hbar\hat{\sigma}_z/2$ the spin operator. This means, for the wave function spinor $\psi_k^{\uparrow\downarrow}$ in a spin spiral background we get 
\begin{align}
    \psi_k^{\uparrow\downarrow}(z+a)=e^{ia(k\hat{1}-q\hat{s}_z/\hbar)}\psi_k^{\uparrow\downarrow}(z).
\end{align}
This corresponds to a generalized Bloch theorem~\cite{sandratskii1998noncollinear}.
Compared to the non-magnetic case, we have replaced
\begin{align}
    k\rightarrow k\hat{1}-q\hat{s}_z/\hbar.
\end{align}
If the bands were spin polarized, we would have a spin-dependent momentum
$\tilde{k}^{\uparrow\downarrow}=k\mp q/2$.
Formerly degenerate spin up and down states are shifted by $\pm q/2$ in reciprocal space [cf. the dashed bands in Fig.~\ref{fig:geometry}(c)]. From this relation it is intuitive to understand that a spin spiral gives rise to a chirality-dependent p-wave magnetism that is characterized by the spin-momentum locking relation
$s_z(k)=-s_z(-k)$
for all bands.
While this result remains true throughout this paper, the actual analysis is more involved because the bands are not actually fully spin polarized: By rotating the local spin frame of reference, the magnetic moments of the spin spiral effectively act as a ferromagnet pointing along an in-plane direction [see Fig.~\ref{fig:geometry}(b)]. This leads to a hybridization of the two shifted bands [dashed lines in Fig.~\ref{fig:geometry}(c)]. A gap opens and the spin states are hybridized [cf. the solid bands in Fig.~\ref{fig:geometry}(c)]. The simplified interpretation based on spin-polarized bands only applies to the case of weak coupling of conduction electron spins to the spiral magnetic texture. 

Depending on the location of the Fermi level, an interesting scenario arises for the Edelstein effect, i.\,e., the non-equilibrium magnetic moment that arises when an electric field is applied along the spiral. If the Fermi level is located in the opened band gap [lower part of Fig.~\ref{fig:geometry}(d)] only one band contributes; one state for positive $k$ and one state with the opposite spin at opposite $k$. When an electric field is applied, the states get shifted by $\Delta k$ and a non-equilibrium magnetic moment arises that is proportional to the shaded volume. Since the two states with opposite spin also have opposite group velocities, their contribution to the Edelstein effect adds up resulting in a rather large magnetic response. When the Fermi level is located elsewhere [upper part of Fig.~\ref{fig:geometry}(d)] both bands contribute and the contributions of the two bands with almost opposite spins almost compensate. This compensated scenario is akin to a Rashba model for which the Edelstein effect is comparably small due to the compensation of the two bands of a spin-split pair of bands~\cite{johansson2021spin}. In both scenarios, for the opposite spiral chirality $q<0$, the spin textures are exactly reversed compared to what is shown in Fig.~\ref{fig:geometry}(d) and therefore the Edelstein response is opposite, i.\,e., chirality dependent. 

\begin{figure}[t!]
    \centering
    \includegraphics[width=\columnwidth]{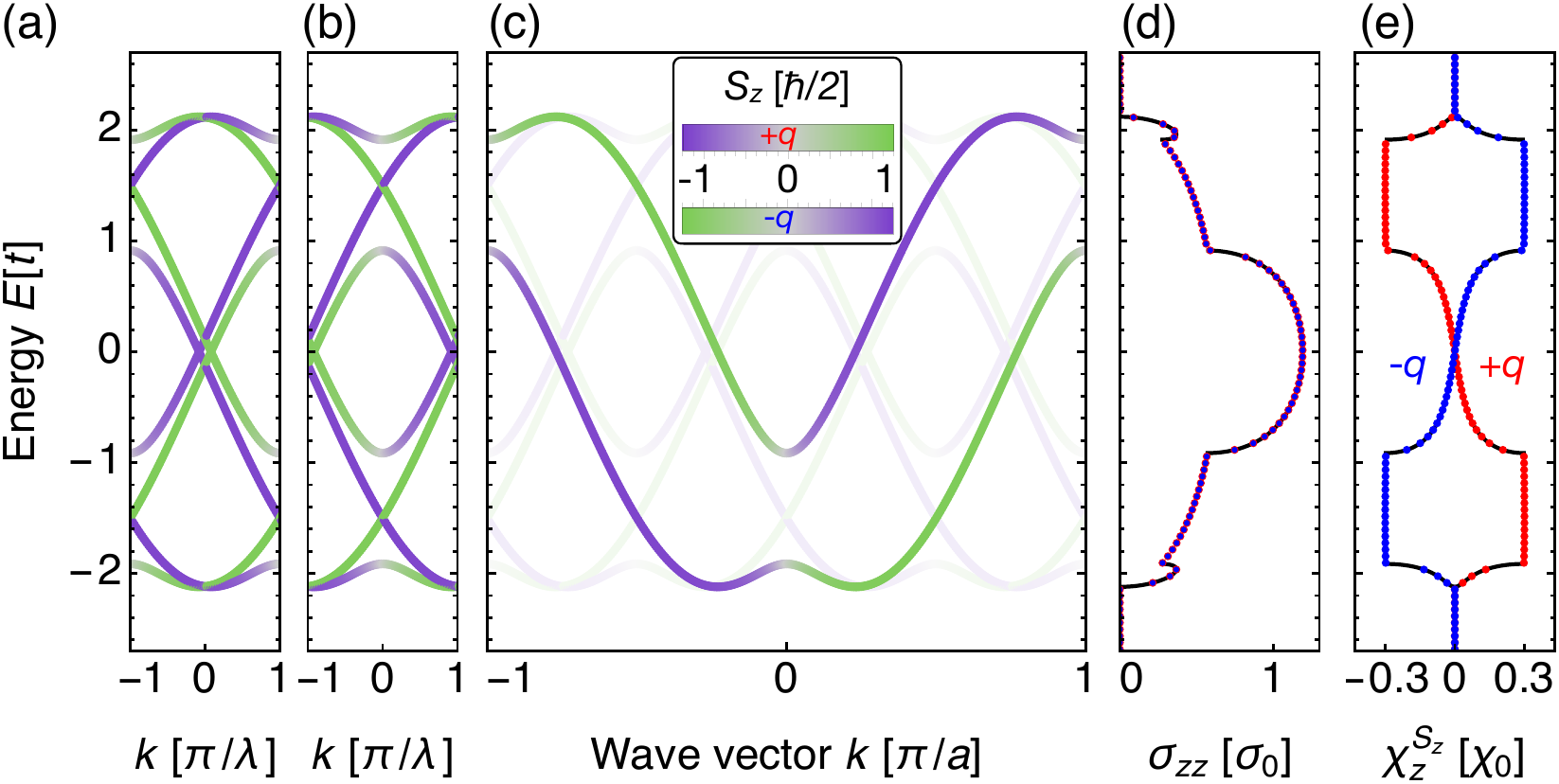}
    \caption{Chirality-dependent properties of a spin spiral. (a) Numerical calculation of the band structure with $s_z(k)$ as the color scale. (b) After applying a unitary transformation, the band structure and spin texture are shifted by $\pm q/2$ as a function of the effectively shifted wave vector. (c) Band structure of the analytical two-band model reproducing the transformed spectrum. The background shows the transformed numerical band structure of panel (b) in the extended Brillouin zone corresponding to the structural unit cell. (d) Longitudinal conductivity and (e) spin Edelstein susceptibility as a function of energy. The color of the data points indicates the chirality. The black curve in the background corresponds to the analytical results. All calculations are performed for $q=\pm\pi/(2a)$ and $J=t/2$. The units are $\chi_0=\frac{e^2g_s\tau a}{m_e}$ and $\sigma_0=\frac{e^2\tau a|t|}{\hbar^2}$.}
    \label{fig:transformation}
\end{figure}

\paragraph*{Exact tight-binding description.}
For a quantitative understanding, we apply this idea to an actual $sd$-Hamiltonian~\cite{ohgushi2000spin}
\begin{align}
    H =-t \sum_{\langle ij\rangle} c_i^\dagger c_j-J \sum_i c_i^\dagger(\vec{m}_i\cdot \boldsymbol\sigma)c_i.\label{eq:ham}
\end{align}
Here, $c_i^\dagger$ and $c_i$ are the creation and annihilation operators of a state at site $i$. The first term describes the hopping quantified by $t$. The second term is the Hund's coupling quantified by $J$: The conduction electrons ($s$ states) interact with the magnetic moments $\{\vec{m}_i\}$ that form the spin spiral based on occupied states that have a lower energy ($d$ states). $\boldsymbol\sigma$ is the vector of Pauli matrices. 

The eigenvalues of $H$ correspond to the band structure $E(k)$ and the eigenvectors determine the spin texture $\vec{s}(k)$. The results for $N=4$ atoms in the unit cell, i.e. $q=\pm\pi/(2a)$, are shown in Fig.~\ref{fig:transformation}(a). While the band structure is chirality independent, the spin texture reverses sign and exhibits p-wave magnetism as predicted before.

The main result of this paper is that we can solve this model fully analytically in perfect agreement with the numerical calculations.
The trick is to gauge away the non-collinearity of the spin spiral, so that the Hund's coupling term in the Hamiltonian becomes trivial. As above, we do this by rotating the spin frame of reference such that all magnetic moments point along $x$. For the Edelstein response, we are interested in the $z$ component of the spin of the electrons, which remains invariant by this transformation. 
The unitary operator that transforms the electron operators $c_i = U_i \psi_i$ is a rotation around the $z$ axis
\begin{align}
    U_i =
e^{- i \frac{q z_i}{2}\sigma_z }.
\end{align}
While the Hund's term becomes trivial, $-J\sum_i \psi_i^\dagger \sigma_x\psi_i$,
the hopping term for nearest neighbors becomes $-t\sum_{i,i\pm1} \psi_i^\dagger e^{\mp i\frac{qa}{2}\hat{\sigma}_z} \psi_{i\pm1}$.
The band structure as a function of the transformed momentum is like the one that was shown before, just shifted by $\pm q/2$ [see Fig.~\ref{fig:transformation}(b) and compare with panel (a)].  For this shift, the sign is not important because $q$ is the size of the Brillouin zone. In both cases, the transformation shifts the $\Gamma$ point to the edge of the Brillouin zone.

After this transformation, each atom has the same hopping terms to its neighbors and the same magnetic texture, pointing along $x$. Therefore, the system can be described by the structural single-atomic unit cell
\begin{align}
    H_\mathrm{eff}=-2t\cos \left(ka - \frac{qa}{2}\hat{\sigma}_z\right)-J\hat{\sigma}_x.
\end{align}
Remarkably, this effective two-band model has no restrictions for $q$. It even describes incommensurate spin spirals that were not possible to describe by the initial $2N\times 2N$ Hamiltonian.

The $2\times2$ effective Hamiltonian matrix can be diagonalized analytically. The band structure
\begin{align}
    E_\pm(k)=-&2t\cos(ka)\cos\left(\frac{qa}{2}\right)\\
    \pm&\sqrt{4t^2\sin^2(ka)\sin^2\left(\frac{qa}{2}\right)+J^2}\notag
\end{align}
is a single band pair in the larger Brillouin zone corresponding to the smaller unit cell. The value of $J$ determines the energy splitting at $k=0$ and $q$ determines the splitting of the bands in $k$ direction.
The result [Fig.~\ref{fig:transformation}(c)] is equivalent to the band structures presented before. By folding back the two bands into the smaller Brillouin zone, corresponding to the larger magnetic unit cell, the band structure from Fig.~\ref{fig:transformation}(b) can be recovered.

The eigenvectors determine the spin texture
\begin{align}
    s_{\pm,z}(k)=\mp \frac{\hbar}{2}\frac{2|t| \sin(ka)\sin\left(\frac{qa}{2}\right)}{\sqrt{4t^2\sin^2(ka)\sin^2\left(\frac{qa}{2}\right)+J^2}}
\end{align}
confirming the chirality-dependent p-wave magnetism $s_{\pm,z}(k)=-s_{\pm,z}(-k)$. The sign is chirality dependent. The results are plotted as a color code in Fig.~\ref{fig:transformation}(a-c).

\paragraph*{Electric field responses.}
Beyond this main result of a chirality-dependent p-wave spin texture, we can even analytically calculate the electric field responses.
When an electric field $\vec{\varepsilon}=\varepsilon_z\vec{e}_z$ is applied along the spin spiral, a non-equilibrium charge current $j_z$ and a non-equilibrium magnetic moment $m_z$ (here normalized to one unit cell) are generated, respectively. The corresponding linear field responses are quantified by the longitudinal wire conductivity $\sigma_{zz}$ and the Edelstein susceptibility $\chi_z^{s_z}$
\begin{align}
    j_z=\sigma_{zz}\varepsilon_z,\quad\quad m_z=\chi_{z}^{s_z}\varepsilon_z.
\end{align}

\begin{figure}[t!]
    \centering
    \includegraphics[width=\columnwidth]{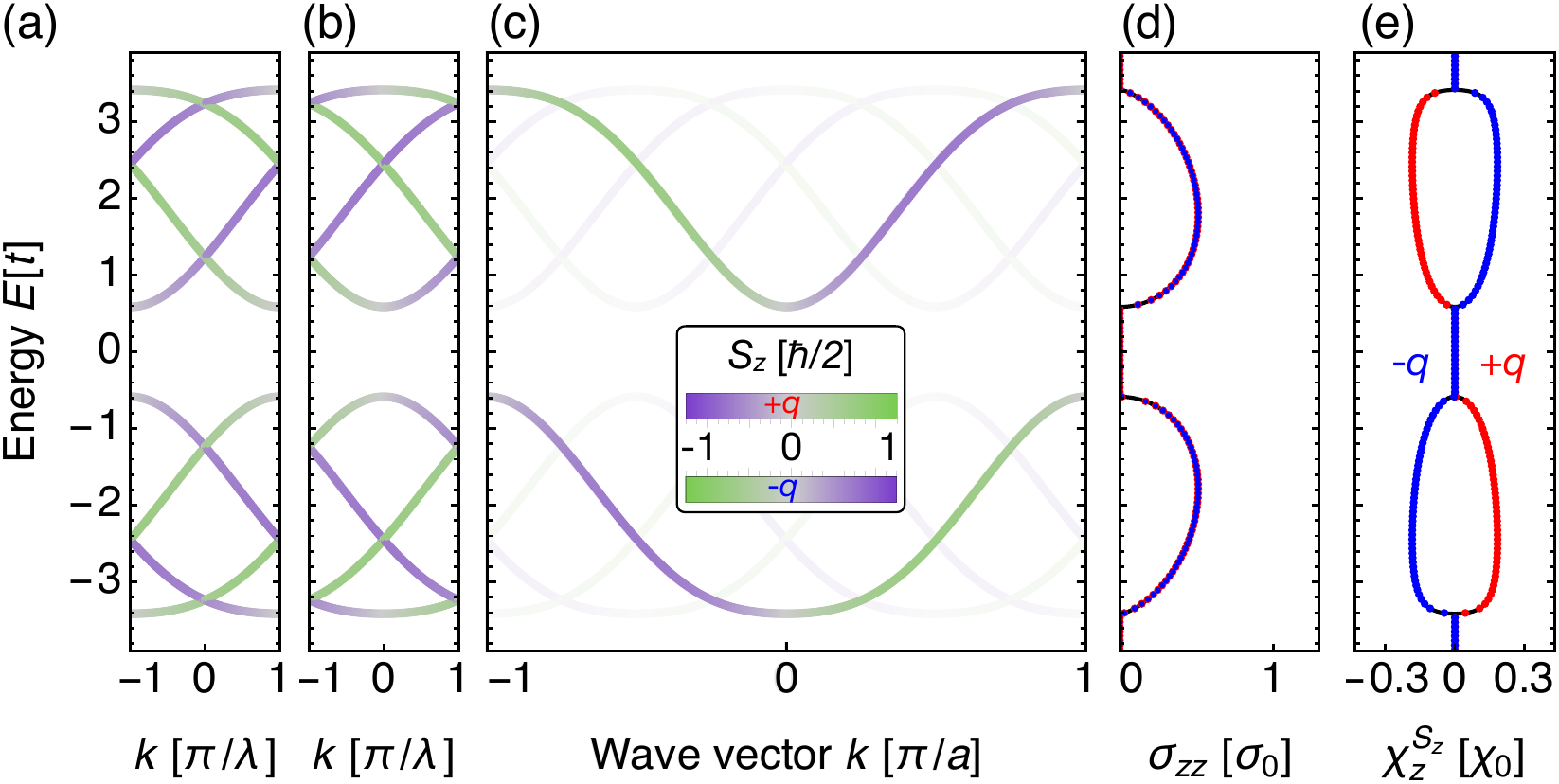}
    \caption{Chirality-dependent properties of a spin spiral in the strong-coupling limit $J=2t$. (a-c) Band structures with spin texture as in Fig. \ref{fig:transformation}. (d) Longitudinal conductivity and (e) Spin Edelstein susceptibility as a function of energy.}
    \label{fig:strong}
\end{figure}

Based on the spin texture and group velocity $v_{n,z}=\frac{1}{\hbar}\frac{\partial E_n(k)}{\partial k}$
we can calculate the chirality-dependent $\chi_{z}^{s_z}$ analytically using Boltzmann transport theory and assuming a constant relaxation time $\tau$~\cite{johansson2024theory}
\begin{align}
    \chi_{z}^{s_z}=\frac{e^2g_s}{2m_e}\tau\sum_{n,k} v_{n,z}(k)\cdot s_{n,z}(k)\cdot\delta(E_n(k)-E).
\end{align}
$m_e$ is the electron mass, $e$ the electron charge and $g_s$ the Landé factor. 
Per energy, there are up to four Fermi wave vectors whose individual contributions to the response tensor elements have to be added up. We exploit symmetry to write
\begin{widetext}
\begin{align}
    \chi_z^{s_z}=-\frac{e^2g_s\tau}{m_e |q|} \sin\left(\frac{qa}{2}\right)\cdot\sum_{\nu=\pm1}\nu|t|\left(4t^2\sin^2\left(\frac{qa}{2}\right)+\frac{4t^2J^2}{4t^2-\left(\sqrt{J^2+(4t^2-E_\mathrm{F}^2)\sin^2\left(\frac{qa}{2}\right)}-\cos\left(\frac{qa}{2}\right)E_\mathrm{F}\nu\right)^2}\right)^{-\nicefrac{1}{2}}. \label{eq:anaedelstein}
\end{align}
\end{widetext}
More details on the derivation can be found in the Supplementary Material where we also derive the analytical expression of the charge conductivity.

The Edelstein susceptibility is an odd function in $q$ which means that it is chirality-dependent. By contrast, the longitudinal conductivity is chirality-independent. The expressions have been plotted as functions of the Fermi level in Figs.~\ref{fig:transformation}(d,e) (black curves) and agree perfectly with numerically calculated data points (red and blue for opposite chiralities).

The longitudinal conductivity [Fig.~\ref{fig:transformation}(d)] is largest near $E\sim 0$ where the group velocity is large and where both bands contribute. The Edelstein susceptibility [Fig.~\ref{fig:transformation}(e)], however, is largest for energies where only two states with opposite spin and opposite group velocity contribute. This corresponds to the lower schematic in Fig.~\ref{fig:geometry}(d). For energies near zero, the Edelstein susceptibility is compensated due to opposite contributions from the two bands with opposite spins.

\paragraph*{Strong-coupling limit.}
For a large Hund's coupling $J>t$, we see two blocks of bands [see Fig.~\ref{fig:strong}(a)] that correspond to parallel and antiparallel spin alignment with the spin spiral, respectively. Again, we see a chirality-dependent p-wave magnetism, chirality-independent longitudinal conductivities [Fig.~\ref{fig:strong}(d)] and a chirality-dependent Edelstein effect [Fig.~\ref{fig:strong}(e)] whose sign is opposite when comparing the upper and lower block of bands. For this set of parameters, we always have the scenario for which only two states contribute that have opposite group velocities and spin expectation values [cf. lower sketch in Fig.~\ref{fig:geometry}(d)]. Therefore, one might expect a large Edelstein effect due to additive contributions. However, note that the spin spiral moments are oriented in-plane, so for large coupling $J>t$ the magnitude of $s_z(k)$ is reduced overall. In fact, the maximum value of $\chi_z^{s_z}$ is smaller than the maximum for the case $J<t$, discussed before.

\begin{figure}[t!]
    \centering
    \includegraphics[width=\columnwidth]{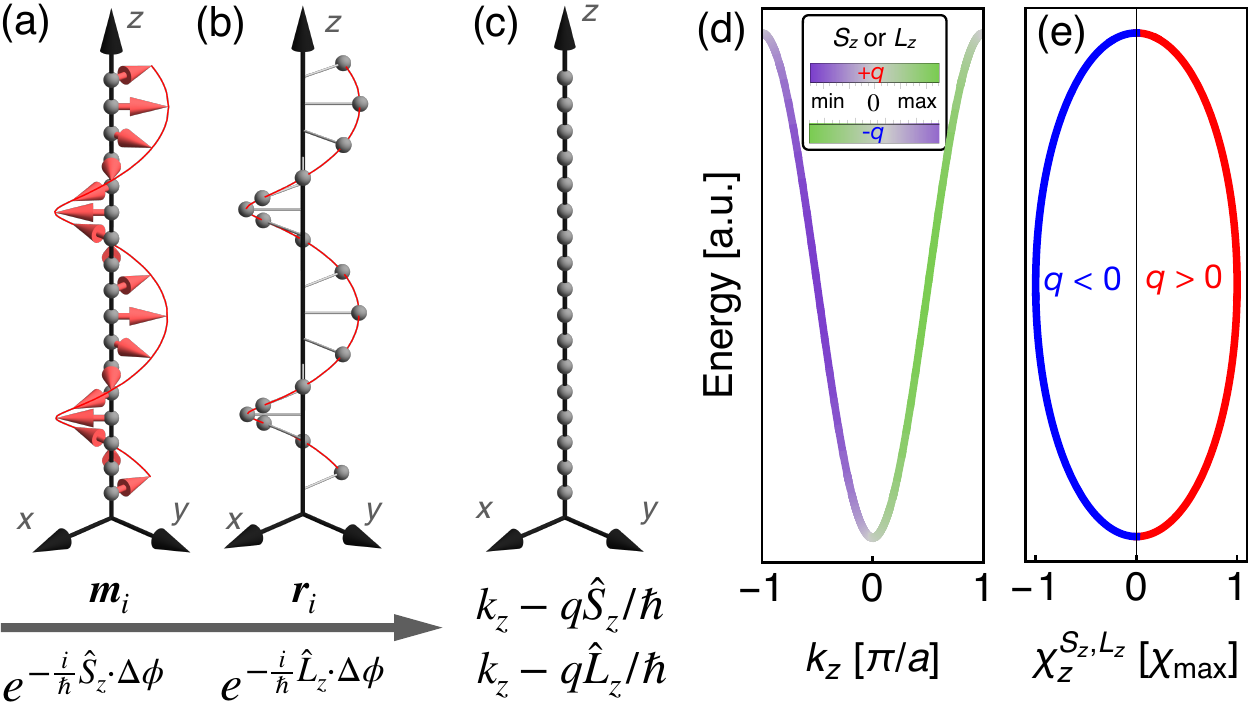}
    \caption{Comparison of chiral systems: Spin-dependent physics in a spin spiral versus orbital-dependent physics in a non-magnetic helix. (a) Spin spiral, (b) non-magnetic helix, (c) transformed systems: a ferromagnetic or non-magnetic linear chain. The transformation is a rotation of the local spin or position vector, respectively and can be expressed by a unitary operator that contains the spin or orbital angular momentum operator, respectively, as indicated. For the spin spiral, the strong coupling limit $J\gg t$ has been considered and only the states with parallel spin alignment are analyzed; the system acts effectively like a non-magnet shifted in energy. (d) The band structure is a single cosine band in both cases. For the spiral, only the energetically lower band is analyzed. p-wave magnetic spin and orbital textures emerge (see color scale). (e) Spin and orbital Edelstein susceptibilities that are chirality dependent in both scenarios.}
    \label{fig:comparison}
\end{figure}

In the limit $J\gg t$, we observe two well separated bands shifted by $\pm J$ respectively, corresponding to parallel and anti-parallel spin alignment with the spin spiral. The band structure of each block essentially becomes a shifted cosine band $E_\pm(k)=-2\tilde{t}\cos(ka)\pm J$
as in the non-magnetic case ($J=0$), with $\tilde{t}=t\cos\left(\frac{qa}{2}\right)$. The spin texture becomes a sine function 
$s_{z,\pm}(k)= \mp \frac{\hbar}{2}\cdot2t \sin(ka)\sin\left(\frac{qa}{2}\right)/J$
and therefore proportional to the band velocity. However, note that the magnitude converges to zero with $J\rightarrow\infty$ because the conduction electron spins align with the in-plane spin spiral. The result is presented in Fig.~\ref{fig:comparison}(d). The considered electric field responses become proportional to each other
\begin{align}
    \chi_z^{s_z}(E)\propto \pm\sigma_{zz}(E)\propto \pm\sqrt{1-\left(\frac{E\pm J}{2\tilde{t}}\right)^2}
\end{align}
as shown in Fig.~\ref{fig:comparison}(e). Note that for spin spirals with low $q$, this limit is reached already for smaller values of $J$.

\paragraph*{Relation to orbital-dependent physics in a helix.}
Interestingly, these electric field responses in the strong coupling limit follow the same functional dependence as the orbital Edelstein susceptibility $\chi_z^{L_z}(E)\propto\sigma_{zz}(E)\propto\sqrt{1-\left(\frac{E}{2t}\right)^2}$ that we have previously predicted in a nonmagnetic helix \cite{gobel2025chirality}. Note that the orbital Edelstein effect~\cite{go2017toward,yoda2018orbital,johansson2021spin,gao2025nonlocal} characterizes the generation of a non-equilibrium orbital magnetic moment when the electric field is applied along the helix.

The reason is the fundamental relation between a helix and a spin spiral as presented in Fig.~\ref{fig:comparison}. Both are quasi-one-dimensional chiral systems -- for the spin spiral the chirality enters via the magnetic moments and for the helix it enters via the position of the atoms. For the helix we can formulate a generalized Bloch theorem as well and apply a similar unitary transformation as presented before for the spin spiral
\begin{align}
    \hat{R}_z=e^{-i\phi\hat{S}_z/\hbar}\quad\rightarrow\quad \hat{R}_z=e^{-i\phi\hat{L}_z/\hbar}.
\end{align}
The rotation operators are equivalent; just the spin operator has to be replaced by the orbital angular momentum operator. Note, that the definition in a periodic system is non-trivial and relies on the modern theory of orbital magnetization~\cite{chang1996berry,xiao2005berry,thonhauser2005orbital}.

Instead of a rotation in spin space, as discussed before, we now rotate in real space to map the helix to an ordinary chain [cf. Figs.~\ref{fig:comparison}(a-c)]. 
The result is a locking of orbital-angular-momentum and crystal momentum resembling p-wave orbital magnetism
$L_z(k)=-L_z(-k)$.
The spin-dependent physics in a strongly coupled spin spiral is equivalent to the orbital-dependent physics in a helix; see Fig.~\ref{fig:comparison}(d,e). An analytical solution of orbital p-wave magnetism and chirality-dependent orbital Edelstein effect of the helix is provided in detail in Ref.~\cite{gobel2025chirality} with an application to the $s$ bands of Tellurium. The orbital p-wave magnetism of these states has recently been confirmed by circular dichroism ARPES~\cite{oh2026observation}.

\paragraph*{Conclusion and outlook.}
A spin spiral gives rise to a chirality-dependent spin-momentum locking, resembling p-wave magnetism, and a chirality-dependent Edelstein effect. In contrast to conventional Rashba physics, the system can be tuned such that there is no partial compensation of the Edelstein effect from spin-split band pairs making the Edelstein effect considerable and a direct consequence of the system's chirality. We can derive all the presented results fully analytically based on an effective two-band description. The generalized Bloch theorem allows for an intuitive understanding based on a chirality- and spin-dependent shift of the wave vector. Our results show that non-collinear magnetism -- even in its simplest form -- gives rise to considerable field responses. The spin spiral is a fundamental and simple archetype of p-wave magnetism that does not rely on spin-orbit coupling. These results identify spin spirals as promising candidates for chirality-dependent spin filtering and provide a mechanism potentially relevant for chirality-induced spin selectivity (CISS)~\cite{ray1999asymmetric,naaman2012chiral,evers2022theory} demonstrating that it can, in principle, emerge entirely from magnetic chirality without spin-orbit coupling.

Our research motivates the analysis of spin-momentum locking in further non-collinear textures such as kagome antiferromagnets~\cite{nakatsuji2015large,nayak2016large,busch2023orbital} or magnetic textures in higher-dimensions, such as magnetic skyrmions~\cite{muhlbauer2009skyrmion,nagaosa2013topological,gobel2024topological} or hopfions~\cite{zheng2023hopfion,gobel2025hopfion}, and their relation to the predicted and observed charge, spin and orbital transport phenomena. 
Furthermore, the equivalence of the spin-dependent physics in a spin spiral and the orbital analogue in a non-magnetic helix~\cite{gobel2025chirality} is not only fascinating from a fundamental point of view. In principle, both chiralities can coexist. For example, a chiral crystal structure gives rise to a Dzyaloshinskii-Moriya interaction (DMI) and therefore also chiral spin textures, as in B20 materials~\cite{muhlbauer2009skyrmion,grigoriev2013chiral,koretsune2015control}. As the next step, we consider the interplay of spin and orbital p-wave magnetism in such systems, especially when spin-orbit interaction is considered~\cite{saunderson2026coupled}.


\paragraph*{Acknowledgements ---}
This work was supported by the EIC Pathfinder OPEN grant 101129641 ``Orbital Engineering for Innovative Electronics'' and by Deutsche Forschungsgemeinschaft (DFG): Project No. 328545488 – CRC/TRR 227, Project No. B12 and by the German Excellence Strategy –EXC3112/1 –533767171 (Center for Chiral Electronics).



\bibliography{short,MyLibrary}

@article{evers2022theory,
  title={Theory of chirality induced spin selectivity: Progress and challenges},
  author={Evers, Ferdinand and Aharony, Amnon and Bar-Gill, Nir and Entin-Wohlman, Ora and Hedeg{\aa}rd, Per and Hod, Oded and Jelinek, Pavel and Kamieniarz, Grzegorz and Lemeshko, Mikhail and Michaeli, Karen and others},
  journal={Advanced {M}aterials},
  volume={34},
  number={13},
  pages={2106629},
  year={2022},
  publisher={Wiley Online Library},
  doi={10.1002/adma.202106629}
}

@article{naaman2012chiral,
  title={Chiral-induced spin selectivity effect},
  author={Naaman, Ron and Waldeck, David H},
  journal={The journal of physical chemistry letters},
  volume={3},
  number={16},
  pages={2178--2187},
  year={2012},
  publisher={ACS Publications},
  doi={10.1021/jz300793y}
}

@article{gao2025nonlocal,
  title={Nonlocal electrical detection of reciprocal orbital Edelstein effect},
  author={Gao, Weiguang and Liao, Liyang and Isshiki, Hironari and Budai, Nico and Kim, Junyeon and Lee, Hyun-Woo and Lee, Kyung-Jin and Go, Dongwook and Mokrousov, Yuriy and Miwa, Shinji and others},
  journal=NatureComm,
  volume={16},
  number={1},
  pages={6380},
  year={2025},
  publisher={Nature Publishing Group UK London},
  doi={10.1038/s41467-025-61602-7}
}

@article{oh2026observation,
  title={Observation of spin-free interatomic orbital angular momentum in a chiral crystal},
  author={Oh, Dongjin and Hahn, Sungsoo and Pacella, Chiara and Yoo, Junseo and Rubio, Angel and Di Sante, Domenico and Kim, Changyoung},
  journal={arXiv preprint: 2605.21124},
  year={2026}
}

@article{sandratskii1998noncollinear,
  title={Noncollinear magnetism in itinerant-electron systems: theory and applications},
  author={Sandratskii, LM},
  journal={Advances in {P}hysics},
  volume={47},
  number={1},
  pages={91--160},
  year={1998},
  publisher={Taylor \& Francis},
  doi={10.1080/000187398243573}
}

@article{koretsune2015control,
  title={Control of Dzyaloshinskii-Moriya interaction in Mn1- x Fe x Ge: a first-principles study},
  author={Koretsune, Takashi and Nagaosa, Naoto and Arita, Ryotaro},
  journal={Scientific {R}eports},
  volume={5},
  number={1},
  pages={13302},
  year={2015},
  publisher={Nature Publishing Group UK London},
  doi={10.1038/srep13302}
}

@article{grigoriev2013chiral,
  title={Chiral Properties of Structure and Magnetism in {M}n$_{1-x}${F}e$_x${G}e Compounds: When the Left and the Right are Fighting, {W}ho Wins?},
  author={Grigoriev, SV and Potapova, NM and Siegfried, S-A and Dyadkin, VA and Moskvin, EV and Dmitriev, V and Menzel, D and Dewhurst, CD and Chernyshov, D and Sadykov, RA and others},
  journal=PRL,
  volume={110},
  number={20},
  pages={207201},
  year={2013},
  publisher={APS},
  doi={10.1103/PhysRevLett.110.207201}
}

@article{chakraborty2025highly,
  title={Highly efficient non-relativistic {E}delstein effect in nodal p-wave magnets},
  author={Chakraborty, Atasi and Birk Hellenes, Anna and Jaeschke-Ubiergo, Rodrigo and Jungwirth, Tom{\'a}s and {\v{S}}mejkal, Libor and Sinova, Jairo},
  journal={Nature Communications},
  volume={16},
  number={1},
  pages={7270},
  year={2025},
  publisher={Nature Publishing Group UK London},
  doi={10.1038/s41467-025-62516-0}
}

@article{brekke2024minimal,
  title={Minimal models and transport properties of unconventional p-wave magnets},
  author={Brekke, Bj{\o}rnulf and Sukhachov, Pavlo and Giil, Hans Gl{\o}ckner and Brataas, Arne and Linder, Jacob},
  journal=PRL,
  volume={133},
  number={23},
  pages={236703},
  year={2024},
  publisher={APS},
  doi={10.1103/PhysRevLett.133.236703}
}

@article{song2025electrical,
  title={Electrical switching of a p-wave magnet},
  author={Song, Qian and Stavri{\'c}, Srdjan and Barone, Paolo and Droghetti, Andrea and Antonenko, Daniil S and Venderbos, J{\"o}rn WF and Occhialini, Connor A and Ilyas, Batyr and Erge{\c{c}}en, Emre and Gedik, Nuh and others},
  journal=Nature,
  volume={642},
  number={8066},
  pages={64--70},
  year={2025},
  publisher={Nature Publishing Group UK London},
  doi={10.1038/s41586-025-09034-7}
}

@article{yoshimori1959new,
  title={A new type of antiferromagnetic structure in the rutile type crystal},
  author={Yoshimori, Akio},
  journal={Journal of the Physical Society of Japan},
  volume={14},
  number={6},
  pages={807--821},
  year={1959},
  publisher={The Physical Society of Japan},
  doi={10.1143/JPSJ.14.807}
}

@article{yamada2025metallic,
  title={A metallic p-wave magnet with commensurate spin helix},
  author={Yamada, Rinsuke and Birch, Max T and Baral, Priya R and Okumura, Shun and Nakano, Ryota and Gao, Shang and Ezawa, Motohiko and Nomoto, Takuya and Masell, Jan and Ishihara, Yuki and others},
  journal=Nature,
  volume={646},
  number={8086},
  pages={837--842},
  year={2025},
  publisher={Nature Publishing Group UK London},
  doi={10.1038/s41586-025-09633-4}
}

@article{hellenes2023p,
  title={P-wave magnets},
  author={Hellenes, Anna Birk and Jungwirth, Tom{\'a}{\v{s}} and Jaeschke-Ubiergo, Rodrigo and Chakraborty, Atasi and Sinova, Jairo and {\v{S}}mejkal, Libor},
  journal={arXiv preprint: 2309.01607},
  year={2023}
}

@article{vsmejkal2022emerging,
  title={Emerging research landscape of altermagnetism},
  author={{\v{S}}mejkal, Libor and Sinova, Jairo and Jungwirth, Tomas},
  journal=PRX,
  volume={12},
  number={4},
  pages={040501},
  year={2022},
  publisher={APS},
  doi={10.1103/PhysRevX.12.040501}
}

@article{fedchenko2024observation,
  title={Observation of time-reversal symmetry breaking in the band structure of altermagnetic {R}u{O}$_2$},
  author={Fedchenko, Olena and Min{\'a}r, Jan and Akashdeep, Akashdeep and D’souza, Sunil Wilfred and Vasilyev, Dmitry and Tkach, Olena and Odenbreit, Lukas and Nguyen, Quynh and Kutnyakhov, Dmytro and Wind, Nils and others},
  journal={Science Advances},
  volume={10},
  number={5},
  pages={eadj4883},
  year={2024},
  publisher={American Association for the Advancement of Science},
  doi={10.1126/sciadv.adj4883}
}

@article{smejkal2022beyond,
  author  = {Libor {\v{S}}mejkal and Jairo Sinova and Tom{\'a}{\v{s}} Jungwirth},
  title   = {Beyond Conventional Ferromagnetism and Antiferromagnetism: A Phase with Nonrelativistic Spin and Crystal Rotation Symmetry},
  journal = PRX,
  volume  = {12},
  pages   = {031042},
  year    = {2022},
  doi     = {10.1103/PhysRevX.12.031042}
}

@article{amin2024nanoscale,
  title={Nanoscale imaging and control of altermagnetism in MnTe},
  author={Amin, OJ and Dal Din, A and Golias, E and Niu, Y and Zakharov, A and Fromage, SC and Fields, CJB and Heywood, SL and Cousins, RB and Maccherozzi, F and others},
  journal=Nature,
  volume={636},
  number={8042},
  pages={348--353},
  year={2024},
  publisher={Nature Publishing Group UK London},
  doi={10.1038/s41586-024-08234-x}
}

@article{saunderson2026coupled,
  title={Coupled Spin-Orbital $ p $-Wave Magnetism via Structural and Magnetic Chirality},
  author={Saunderson, Tom G and G{\"o}bel, B{\"o}rge and {\c{S}}a{\c{s}}{\i}o{\u{g}}lu, Ersoy and Lounis, Samir},
  journal={arXiv preprint:2607.02378},
  year={2026}
}

@article{gobel2025hopfion,
  title={Three-dimensional topological orbital {H}all effect caused by magnetic hopfions},
  author={Göbel, Börge and Lounis, Samir},
  journal=PRB,
  volume={112},
  pages={134426},
  year={2025},
  publisher={APS},
  doi={10.1103/sc9g-l9by}
}

@article{zheng2023hopfion,
  title={Hopfion rings in a cubic chiral magnet},
  author={Zheng, Fengshan and Kiselev, Nikolai S and Rybakov, Filipp N and Yang, Luyan and Shi, Wen and Bl{\"u}gel, Stefan and Dunin-Borkowski, Rafal E},
  journal={Nature},
  volume={623},
  number={7988},
  pages={718--723},
  year={2023},
  publisher={Nature Publishing Group UK London},
  doi={10.1038/s41586-023-06658-5}
}

@article{gobel2025chirality2,
  title={Chirality-induced selectivity of angular momentum by orbital {E}delstein effect in carbon nanotubes},
  author={G{\"o}bel, B{\"o}rge and Mertig, Ingrid and Lounis, Samir},
  journal={Communications Physics},
  volume={8},
  number={1},
  pages={395},
  year={2025},
  publisher={Nature Publishing Group UK London},
  doi={10.1038/s42005-025-02331-7}
}

@article{gobel2025chirality,
  title={Chirality-induced orbital {E}delstein effect in an analytically solvable model},
  author={G{\"o}bel, B{\"o}rge and Schimpf, Lennart and Mertig, Ingrid},
  journal=PRR,
  volume={7},
  number={3},
  pages={033180},
  year={2025},
  doi={10.1103/vpjm-ntbh}
}

@article{yoda2018orbital,
  title={Orbital {E}delstein effect as a condensed-matter analog of solenoids},
  author={Yoda, Taiki and Yokoyama, Takehito and Murakami, Shuichi},
  journal=NanoLett,
  volume={18},
  number={2},
  pages={916--920},
  year={2018},
  publisher={ACS Publications},
  doi={10.1021/acs.nanolett.7b04300}
}

@article{johansson2024theory,
  title={Theory of spin and orbital {E}delstein effects},
  author={Johansson, Annika},
  journal=JPhysCM,
  volume={36},
  number={42},
  pages={423002},
  year={2024},
  publisher={IOP Publishing},
  doi={10.1088/1361-648X/ad5e2b}
}

@article{bychkov1984oscillatory,
  title={Oscillatory effects and the magnetic susceptibility of carriers in inversion layers},
  author={Bychkov, Yu A and Rashba, Emmanuel I},
  journal={Journal of physics C: Solid state physics},
  volume={17},
  number={33},
  pages={6039},
  year={1984},
  publisher={IOP Publishing},
  doi={10.1088/0022-3719/17/33/015}
}

@article{rashba1960properties,
  title={Properties of semiconductors with an extremum loop},
  author={Rashba, EIJSP},
  journal={Sov. Phys.-Solid State},
  volume={2},
  pages={1109},
  year={1960}
}

@article{edelstein1990spin,
  title={Spin polarization of conduction electrons induced by electric current in two-dimensional asymmetric electron systems},
  author={Edelstein, Victor M},
  journal={Solid State Communications},
  volume={73},
  number={3},
  pages={233--235},
  year={1990},
  publisher={Elsevier},
  doi={10.1016/0038-1098(90)90963-C}
}

@article{ray1999asymmetric,
  title={Asymmetric scattering of polarized electrons by organized organic films of chiral molecules},
  author={Ray, K and Ananthavel, SP and Waldeck, DH and Naaman, Ron},
  journal=Science,
  volume={283},
  number={5403},
  pages={814--816},
  year={1999},
  publisher={American Association for the Advancement of Science},
  doi={10.1126/science.283.5403.814}
}

@article{gobel2024topological,
  title={Topological orbital {H}all effect caused by skyrmions and antiferromagnetic skyrmions},
  author={G{\"o}bel, B{\"o}rge and Schimpf, Lennart and Mertig, Ingrid},
  journal={Communications Physics},
  volume={8},
  number={1},
  pages={17},
  year={2025},
  publisher={Nature Publishing Group UK London},
  doi={10.1038/s42005-024-01925-x}
}

@article{gobel2021beyond,
  title={Beyond skyrmions: {R}eview and perspectives of alternative magnetic quasiparticles},
  author={G{\"o}bel, B{\"o}rge and Mertig, Ingrid and Tretiakov, Oleg A},
  journal={Physics Reports},
  volume={895},
  pages={1--28},
  year={2021},
  publisher={Elsevier},
  doi={10.1016/j.physrep.2020.10.001}
}

@article{johansson2021spin,
  title={Spin and orbital {E}delstein effects in a two-dimensional electron gas: {T}heory and application to {SrTiO}$_3$ interfaces},
  author={Johansson, Annika and G{\"o}bel, B{\"o}rge and Henk, J{\"u}rgen and Bibes, Manuel and Mertig, Ingrid},
  journal=PRR,
  volume={3},
  number={1},
  pages={013275},
  year={2021},
  publisher={APS},
  doi={10.1103/PhysRevResearch.3.013275}
}

@article{busch2023orbital,
  title={Orbital {H}all effect and orbital edge states caused by $s$ electrons},
  author={Busch, Oliver and Mertig, Ingrid and G{\"o}bel, B{\"o}rge},
  journal=PRR,
  volume={5},
  number={4},
  pages={043052},
  year={2023},
  publisher={APS},
  doi={10.1103/PhysRevResearch.5.043052}
}

@article{thonhauser2005orbital,
  title={Orbital magnetization in periodic insulators},
  author={Thonhauser, Timo and Ceresoli, Davide and Vanderbilt, David and Resta, Raffaele},
  journal=PRL,
  volume={95},
  number={13},
  pages={137205},
  year={2005},
  publisher={APS},
  doi = {10.1103/PhysRevLett.95.137205},
  url = {https://link.aps.org/doi/10.1103/PhysRevLett.95.137205}
}

@article{xiao2005berry,
  title={Berry phase correction to electron density of states in solids},
  author={Xiao, Di and Shi, Junren and Niu, Qian},
  journal=PRL,
  volume={95},
  number={13},
  pages={137204},
  year={2005},
  publisher={APS},
  doi = {10.1103/PhysRevLett.95.137204},
  url = {https://link.aps.org/doi/10.1103/PhysRevLett.95.137204}
}

@article{chang1996berry,
  title={Berry phase, hyperorbits, and the {H}ofstadter spectrum: Semiclassical dynamics in magnetic {B}loch bands},
  author={Chang, Ming-Che and Niu, Qian},
  journal=PRB,
  volume={53},
  number={11},
  pages={7010},
  year={1996},
  publisher={APS},
  doi = {10.1103/PhysRevB.53.7010},
  url = {https://link.aps.org/doi/10.1103/PhysRevB.53.7010}
}

@article{hasan2010colloquium,
  title={Colloquium: topological insulators},
  author={Hasan, M Zahid and Kane, Charles L},
  journal=RevModPhys,
  volume={82},
  number={4},
  pages={3045},
  year={2010},
  publisher={APS},
  doi={10.1103/RevModPhys.82.3045}
}

@article{ohgushi2000spin,
  title={Spin anisotropy and quantum {H}all effect in the kagom{\'e} lattice: {C}hiral spin state based on a ferromagnet},
  author={Ohgushi, Kenya and Murakami, Shuichi and Nagaosa, Naoto},
  journal=PRB,
  volume={62},
  number={10},
  pages={6065(R)},
  year={2000},
  publisher={APS},
  doi={10.1103/PhysRevB.62.R6065}
}

@article{nagaosa2013topological,
  title={Topological properties and dynamics of magnetic skyrmions},
  author={Nagaosa, Naoto and Tokura, Yoshinori},
  journal=NatureNano,
  volume={8},
  number={12},
  pages={899--911},
  year={2013},
  publisher={Nature Publishing Group},
  doi={10.1038/nnano.2013.243}
}

@article{muhlbauer2009skyrmion,
  title={Skyrmion lattice in a chiral magnet},
  author={M{\"u}hlbauer, S and Binz, B and Jonietz, F and Pfleiderer, C and Rosch, A and Neubauer, A and Georgii, R and B{\"o}ni, P},
  journal={Science},
  volume={323},
  number={5916},
  pages={915--919},
  year={2009},
  publisher={American Association for the Advancement of Science},
  doi={10.1126/science.1166767}
}

@article{nakatsuji2015large,
  title={Large anomalous {H}all effect in a non-collinear antiferromagnet at room temperature},
  author={Nakatsuji, Satoru and Kiyohara, Naoki and Higo, Tomoya},
  journal={Nature},
  volume={527},
  number={7577},
  pages={212},
  year={2015},
  publisher={Nature Publishing Group},
  doi={10.1038/nature15723},
  url={10.1038/nature15723}
}

@article{nayak2016large,
  title={Large anomalous {H}all effect driven by a nonvanishing {B}erry curvature in the noncolinear antiferromagnet {Mn$_3$Ge}},
  author={Nayak, Ajaya K and Fischer, Julia Erika and Sun, Yan and Yan, Binghai and Karel, Julie and Komarek, Alexander C and Shekhar, Chandra and Kumar, Nitesh and Schnelle, Walter and K{\"u}bler, J{\"u}rgen and others},
  journal={Science Advances},
  volume={2},
  number={4},
  pages={e1501870},
  year={2016},
  publisher={American Association for the Advancement of Science},
  doi = {10.1126/sciadv.1501870},
  URL = {https://www.science.org/doi/abs/10.1126/sciadv.1501870}
}

@article{go2017toward,
  title={Toward surface orbitronics: giant orbital magnetism from the orbital {R}ashba effect at the surface of $sp$-metals},
  author={Go, Dongwook and Hanke, Jan-Philipp and Buhl, Patrick M and Freimuth, Frank and Bihlmayer, Gustav and Lee, Hyun-Woo and Mokrousov, Yuriy and Bl{\"u}gel, Stefan},
  journal={Scientific Reports},
  volume={7},
  number={1},
  pages={46742},
  year={2017},
  publisher={Nature Publishing Group UK London},
  doi={10.1038/srep46742},
  url={10.1038/srep46742}
}

@String{JPhysCM = "J. Phys.: Condens.\ Matt."}

@String{NanoLett = "Nano Lett."}

@String{Nature = "Nature"}

@String{NatureNano = "Nature Nanotechnol."}

@String{NatureComm = "Nature Comms."}

@String{PRB = "Phys.\ Rev.\ B"}

@String{PRX = "Phys.\ Rev.\ X"}

@String{PRL = "Phys.\ Rev.\ Lett."}

@String{PRR = "Phys.\ Rev.\ Res."}

@String{RevModPhys = "Rev.\ Mod.\ Phys."}

\end{document}